\documentclass[sn-mathphys]{sn-jnl}
\jyear{2022}%
\theoremstyle{thmstyleone}%
\theoremstyle{thmstyletwo}%

\theoremstyle{thmstylethree}%

\usepackage{adjustbox}
\usepackage[numbers]{natbib}
\usepackage{amsmath}
\usepackage{float}
\usepackage{booktabs,threeparttable,caption}
\usepackage{setspace}
\usepackage{url}
\DeclareUnicodeCharacter{202F}{\,}
\usepackage{newtxtext,newtxmath}
\usepackage{hyperref}

\begin{document}

\title[Classification of sleep stages]{Classification of sleep stages from EEG, EOG and EMG signals by SSNet}

\author*[1]{\fnm{Haifa} \sur{Almutairi}}\email{haifa.almutairi@research.uwa.edu.au}

\author[1]{\fnm{Ghulam Mubashar} \sur{Hassan}}

\author[1]{\fnm{Amitava} \sur{Datta}}

\affil*[1]{\orgdiv{Department of Computer Science and Software Engineering}, \orgname{he University of Western Australia}, \country{Australia}}

\abstract{Classification of sleep stages plays an essential role in diagnosing sleep-related diseases including Sleep Disorder Breathing (SDB) disease. In this study, we propose an end-to-end deep learning architecture, named SSNet, which comprises of two deep learning networks based on Convolutional Neuron Networks (CNN) and Long Short Term Memory (LSTM). Both deep learning networks extract features from the combination of Electrooculogram (EOG), Electroencephalogram (EEG), and Electromyogram (EMG) signals, as each signal has distinct features that help in the classification of sleep stages. The features produced by the two-deep learning networks are concatenated to pass to the fully connected layer for the classification. The performance of our proposed model is evaluated by using two public datasets Sleep-EDF Expanded dataset and ISRUC-Sleep dataset. The accuracy and Kappa coefficient are 96.36\% and 93.40\% respectively, for classifying three classes of sleep stages using Sleep-EDF Expanded dataset. Whereas, the accuracy and Kappa coefficient are 96.57\% and 83.05\% respectively for five classes of sleep stages using Sleep-EDF Expanded dataset. Our model achieves the best performance in classifying sleep stages when compared with the state-of-the-art techniques.}


\maketitle

\section{Introduction}
Sleep is a critical part of human life which helps to maintain good health and quality of life. When a person feels tired after a full night's sleep or fatigue during the day, this can be an indication that the person may be suffering from Sleep Disorders (SD) \cite{wulff2010sleep, ohayon2011epidemiological}. Examples of SD diseases include Sleep Disordered Breathing (SDB) \cite{malhotra2002obstructive}, Periodic Legs Movement (PLM) \cite{ohayon2002prevalence} and Insomnia \cite{morin2012chronic}. A study by Peppard et al.\cite{peppard2013increased} found that about 30\% of the adult population in the United States of America have insomnia. Also, more than 50 million Americans are diagnosed with sleep disorders, and approximately 25 million Americans have SDB \cite{altevogt2006sleep}. An early-stage diagnosis of SD can protect patients from severe diseases including cardiovascular problems, neurocognitive deficits, diabetes, stroke and recurrent heart attacks \cite{almutairi2021detection, malhotra2002obstructive}.\\

Sleep is categorized into five sleep stages according to the guidelines of American Academy of Sleep Medicine (AASM) \cite{khalighi2016isruc}, which are  Wake (W) stage, Non-Rapid Eye Movement stage (NREM) which contains three stages (N1, N2 and N3) and Rapid Eye Movement stage (REM). Normally, people move from W stage to NREM stage followed by REM stage. Each sleep stage's electrical brain activity is recorded by sensors attached to different parts of the body. There are three different types of brain activities: alpha, theta and delta. W stage exhibits an alpha activity which appears in the occipital region. N1 stage is a shallow sleep that characterizes low alpha activity and the occurrence of theta activity \cite{lee2019possible}. The actual sleep starts in N2 stage and a unique waveform is produced which is called \textit{sleep spindle} \cite{lee2019connectivity}. N3 stage is a deep sleep stage characterized by the occurrence of delta wave \cite{garcia2018closed}. Lastly, REM stage is characterized by low-voltage and fast activity in theta waves \cite{nir2013sleep}. Table \ref{frequencies} shows the Characteristic frequency of EEG signals for each sleep stage. The percentages of a normal cycle of sleep stages are: 50–60\% of sleep time spent in the (N1, N2) light sleep stages, 15–20\% of sleep time spent in the (N3) deep sleep stage, 20–25\% of sleep time spent in REM sleep stage, and 5\% or less of the sleep time spent in W sleep stage \cite{penzel2003dynamics}.

\begin{table}[htp]
\begin{center}
		\caption{Characteristic frequency ranges of EEG signals for each sleep stage}			\label{frequencies}
\begin{tabular}{|l|l|}
\hline
\textbf{Sleep stage} & \textbf{Characteristic frequency}                                      \\ \hline
W                    & Alpha (8-12 Hz)                                                          \\ \hline
N1                   & Theta (4-8 Hz)                                                           \\ \hline
N2                   & Spindle (12-15 Hz)                                                       \\ \hline
N3                   & Delta (0.5-4 Hz)                                                         \\ \hline
REM                  & \begin{tabular}[c]{@{}l@{}}Alpha (8-12 Hz)\\ Theta (4-8 Hz)\end{tabular} \\ \hline
\end{tabular}
\end{center}
\end{table}

In a sleep laboratory, polysomnography (PSG) \cite{bloch1997polysomnography} is a standard clinical procedure used for classification of sleep stages. PSG device has multi-sensors to record physiological signals such as Electromyogram (EMG) \cite{shokrollahi2012sleep}, Electrocardiography (ECG) \cite{kesper2012ecg}, Electroencephalogram (EEG) \cite{campbell2009eeg}, and Electrooculogram (EOG) \cite{jammes2008automatic} signals. Sleep experts use manual analysis of physiological signals to classify sleep stages. The drawbacks of manual analysis include time-consuming process, the possibility of human errors in diagnosis, and an inconvenient procedure for patients \cite{collop2002scoring}. Therefore, an automatic procedure for the classification of sleep stages will help for diagnosing SD at hospitals.\\

Due to technologies showing improvement in health care system, machine learning models are developed to evaluate biomedical signals including EEG, ECG, EMG and EOG signals. For example, studies proposed models for different biomedical problems, such as detection of Parkinson’s disease using EEG signals \cite{oh2020deep}, detection of directions of eye movements using EOG signals \cite{banerjee2013classifying}, and detection of atrial fibrillation using ECG signals \cite{andersen2019deep}. Few studies in the literature developed machine learning models for the sleep stage classification. Out of which some studies suggested to extract features from EEG signals and then classifying them using machine learning. They classified 30-second segments into three sleep stages: W, NREM and REM, and five sleep stages: W, N1, N2, N3 and REM. For instance, Hassan et al. \cite{hassan2017automated} proposed Empirical Mode Decomposition (EMD) method and a random under-sampling boosting (RUSBoost) classifier. The segments’ classification accuracy was 94.23\% in the three sleep stage classification and 83.49\% in the five sleep stage classification. Another study by Hassan et al. \cite{hassan2016computer} proposed Complete Ensemble Empirical Mode Decomposition (CEEMD) and Bootstrap Aggregating (Bagging) classifier. The classification accuracy on all segments for the three sleep stage classification was 94.10\%, and for the five sleep stage classification was 90.96\%. Zhu et al. \cite{zhu2014analysis} proposed a graph domain method and a Support vector machine (SVM) to classify the segments into the three and five sleep stages. The accuracy of their model on the classification of the three sleep stages was 92.60\%, and for the five sleep stages was 88.90\%. Sharma et al. \cite{sharma2018accurate} proposed a wavelet filter method and SVM classifier. The segment classification accuracy for the three sleep stages was 93.50\%, and for the five sleep stages was 91.5\%. Satapathy et al. \cite{satapathy2021machine} proposed a model that used statistical features such as mean, variance and skewness. They used a random forest classifier to classify 30-second segments of EEG signals into the five sleep stages and the accuracy was found to be 92.79\%. A study by Rahman et al.\cite{rahman2018sleep} proposed a model that used a discrete wavelet transform method and SVM classifier. The segment classification accuracy for the five sleep stages was found to be 91.70\%.  \\

Recently, researchers proposed Deep Learning (DL) techniques based on Convolutional Neural Networks (CNN) for sleep stage classification. CNN architecture has been very successful in classification \cite{yang2021cnn,xia2018evaluation}, object recognition \cite{7485869} and image segmentation \cite{7803544} problems. Several studies proposed different CNN models to classify 30-second segments into the three and five sleep stages. For instance, Yildirim et al. \cite{yildirim2019deep} proposed a CNN model to extract features from EEG and EOG segments without applying any feature engineering methods. They used 10 layers of 1D-CNN and a fully connected layer. Their model achieved an accuracy of 94.24\% in the three sleep stage classification and 90.98\% in the five sleep stage classification. Nguyen et al. \cite{rajbhandari2021novel} proposed CNN model for the five sleep stage classification. Their architecture contains three layers of 1D-CNN, in which the first 1D-CNN is followed by max-pooling and dropout layers, the second 1D-CNN layer is followed by max-pooling and batch normalization, and the last 1D-CNN is followed by max-pooling and two fully connected layers. Their model achieved an accuracy of 87.67\% in the five sleep stage classification. Similarly, Zhu et al. \cite{zhu2020convolution} proposed a deep learning model based on 1D-CNN and attention mechanism. The segment classification accuracy for the five sleep stages was reported to be 82.80\%.  \\

The existing studies mentioned above share some limitations. Firstly, they involve feature extracting methods which are complicated, time-consuming and computationally complex processes \cite{mirza2019machine}. Secondly, most of the existing works are based on a single channel of EEG signals. We found that some other behaviours such as muscle and eye movements, which record from EMG and EOG signals can also affect sleep irregularities \cite{torres2014electroencephalogram}. We solved these limitations by using a combination of EEG, EMG and EOG signals, which provide distinct features that help us to improve the results.\\


EOG and EMG signals contribute valuable additional sources with EEG signals in the classification of sleep stages. Muscular activities and eye movements appear in EMG and EOG signals during sleep stages. EMG signals show that muscular activities are reduced during NREM sleep stage, whereas muscular activities are lost completely during REM stage. On the other hand, EOG signals show bilateral eye movements during REM stage \cite{torres2014electroencephalogram}. These features from EMG and EOG signals can distinguish between NREM and REM sleep stages. Few studies focused on the classification of 30-second segments of sleep stages based on a combination of EEG, EMG and EOG signals. For instance, Cui et al. \cite{cui2018automatic} proposed a CNN model to extract features from EEG, EMG and EOG segments without applying any feature engineering methods. They used two layers of 2D-CNN followed by max-pooling and the last layer is fully connected. Their model achieved an accuracy of 92\% on a subject wise test set for the five sleep stage classification. Phan et al. \cite{phan2018joint} proposed a Fast Fourier Transform (FFT) method and 2D-CNN model for five sleep stage classification. Their model achieved an accuracy of 83.6\% on a subject wise test set. \\

In this paper, we propose an efficient automatic deep learning model for the classification of the three and five sleep stages. Our proposed model is an end-to-end deep learning model called \textit{SleepStageNet (SSNet)} to classify 30-second segments of the combination of EEG, EMG and EOG signals. Our proposed architecture contains two deep learning networks. The first deep learning network includes a 1D-CNN network to extract time-invariant features from the raw signals. The second deep learning network includes a Long Short Term Memory (LSTM) to extract temporal features from a sequence of the raw signal. A fully connected layer classifies the combined features extracted from both the deep learning networks. SSNet can be used
for the automatic classification of sleep stages at hospitals. It can assist physician experts in analysing PSG signals rather than using manual methods. 
  \\ 


This paper is organized as follows. Section 2 includes data preparation which involves describing two datasets and data distribution. Section 3 describes the proposed SSNet. Section 4 presents the results while Section 5 presents discussion. The conclusion is presented at the end.

\section{Data Preparation}
This section describes two public datasets: ISRUC-Sleep dataset and Sleep-EDF Expanded dataset, and data distribution used in the experiments of this study.

\subsection{ISRUC-Sleep Dataset}
This dataset is collected by Sleep Medicine Centre of Hospital of Coimbra University (CHUC) \cite{khalighi2016isruc}. The total number of PSG recordings is 116 with 11 channels. Each recording includes the following channels:
\begin{itemize}
	\item Six EEG channels with references A1 and A2 (C3-A2), (C4-A1), (F3-A2), (O1-A2), (O2-A1), and (F4-A1) which are placed on the both sides of earlobes.
	\item Two EOG channels (LOC-A2) and (ROC-A1) which are placed on the left and right eye movements.	
	 \item Chine EMG channel (X1) which are placed between the chin and the lower lip.
	\item One channel of ECG signals (X2)
	\item Two EMG channels (X3 and X4) which are placed on the left and right leg movement. 
\end{itemize} 

Each recording was sampled at 200 Hz, and the duration of the recording was around 8 hours. Sleep physicians segmented the recordings into 30-second segments and labelled them according to American Academy of Sleep Medicine (AASM) rules. Each segment was labelled with one of the five sleep stages: W, N1, N2, N3 and REM.\\

ISRUC-Sleep dataset is divided into three subgroups depending on the health status: 
\begin{itemize}
	\item \textbf{Subgroup 1:} the data is recorded from 100 subjects having sleep disorder disease. Each recording belongs to one subject. 
	\item \textbf{Subgroup 2:} the data is recorded from 8 subjects who were under treatment. Two recording sessions are provided per subject.
	\item \textbf{Subgroup 3:} the data is recorded from 10 healthy subjects. Each recording belongs to one subject.
\end{itemize} 

\subsection{Sleep-EDF Expanded Dataset}
Sleep-EDF Expanded dataset is an extended version of Sleep-EDF dataset \cite{goldberger2000physiobank}, which was published publicly on the PhysioBank website in 2013. The total number of PSG recordings is 197. Each recording contains two channels of EEG signals (Fpz-Cz and Pz-Oz electrode locations), one channel of EOG signals (horizontal), and one channel of chin EMG signals. The labels of 30-second segments of each recording are done manually by sleep experts based on AASM guidelines. Each segment is labelled with one of five sleep stages: W, N1, N2, N3, and REM and the sampling rate of each recording is 100 Hz. The dataset is divided into two subgroups \cite{kemp2000analysis}: 
\begin{itemize}
	\item \textbf{Sleep Cassette subgroup (SC*):} it contains 153 recordings. Each two recordings set belongs to one healthy subject. The duration of the two recordings is around 20 hours.    
	
	\item \textbf{Sleep Telemetry subgroup (ST*):} it contains 44 recordings. The duration of each recording is around 9 hours. Each recording belongs to one subject who has mild difficulty in falling asleep. 

\end{itemize}

\subsection{Data Distribution}
We used five channels of EEG, EOG and EMG signals (O1-A2, C3-A2, C4-A1, X1 and LOC-A2) of ISRUC-Sleep dataset as per recommendation of  Cui et al. \cite{cui2018automatic} who suggested that the CNN can extract features from the combination of EEG, EOG and EMG signals to classify the five sleep stages. Each segment has 6000 sampling points and each recording is labelled with a final diagnosis as SDB, Epilepsy \cite{sander1996epidemiology}, Parasomnia \cite{schenck1987rapid}, or other sleep-related disorders. In this study, we selected the recordings that have the final diagnosis of SDB. In Sleep-EDF Expanded dataset, we selected 73 recordings from the (SC*) subgroup and 42 recordings from the (ST*) subgroup randomly. This restricted selection was due to the computational resources limitation including GPU and size of RAM. Each segment has 3000 sampling points. We used all the four available channels of EEG, EOG and EMG signals (Fpz-Cz, Pz-Oz, EOG horizontal and chin EMG).\\

Table \ref{Table 2} shows the details of the selected segments of the two datasets. We randomly selected 25,449 out of 55,824 NREM segments of ISRUC-Sleep dataset and 25,201 out of 77,158 NREM segments of Sleep-EDF Expanded datasets. The reason for decreasing the segments is to prevent overfitting and resolve class imbalance during the classification stages. \\

It is worth mentioning that, we did not use any methods for filtering or noise removal. We only normalised the segments in the two datasets by Z-score as presented in Eq.~(\ref{Eq1}) \cite{mohamad2013standardization}. 
\begin{equation}
z_{score} =\frac { (S - E_S)}{\alpha_S}
\label{Eq1} 
\end{equation}
where $S$ is the segment, $E$ is the mean of the segment and $\alpha$ represents the standard deviation of the segment. \\

SSNet was trained and tested on a system having an Intel (R) Core (TM) 3.6 GHz (i7–7700) processor and 8 GB RAM. We used Python 3.7 version with Keras and Scikit-learn libraries. We used Adam optimization rate of 0.002, cross-entropy loss functions and a batch size of 128. We evaluated the performance of the proposed SSNet with the two datasets. We selected segments randomly for each set: training set has 70\%,  validation set has 15\%, and testing set has 15\% of the datasets. \\

\begin{table*}[]
\begin{center}
		\caption{Distribution of the selected segments of SleepEDFX and ISRUC-Sleep datasets.}	
		 \begin{adjustbox}{width=1.05\columnwidth,center}

\begin{tabular}{|l|l|l|l|l|l|l|l|}
\hline
\textbf{Number of segments}  & \textbf{Dataset} & \textbf{W} & \textbf{N1} & \textbf{N2} & \textbf{N3} & \textbf{R} & \textbf{Total} \\ \hline
Total segments & SleepEDFX        & 25,201      & 10,420       & 52,502       & 14,236       & 21,602      & 123,962        \\ \hline
Selected segments  & SleepEDFX        & 25,201      & 3,207        & 16,748       & 5,246        & 21,602      & 72,000         \\ \hline
Total segments & ISRUC-sleep       & 19,810      & 11,101       & 27,398       & 17,325       & 11,256      & 86,993         \\ \hline
Selected segments  & ISRUC-sleep       & 19,810      & 4,935        & 12,668       & 7,846        & 11,256      & 56,515         \\ \hline
\end{tabular}   \end{adjustbox}
		\label{Table 2}
\end{center}

\end{table*}

\section{The proposed SSNet}
\begin{figure*}[]
\begin{center}
	\includegraphics[scale=.33]{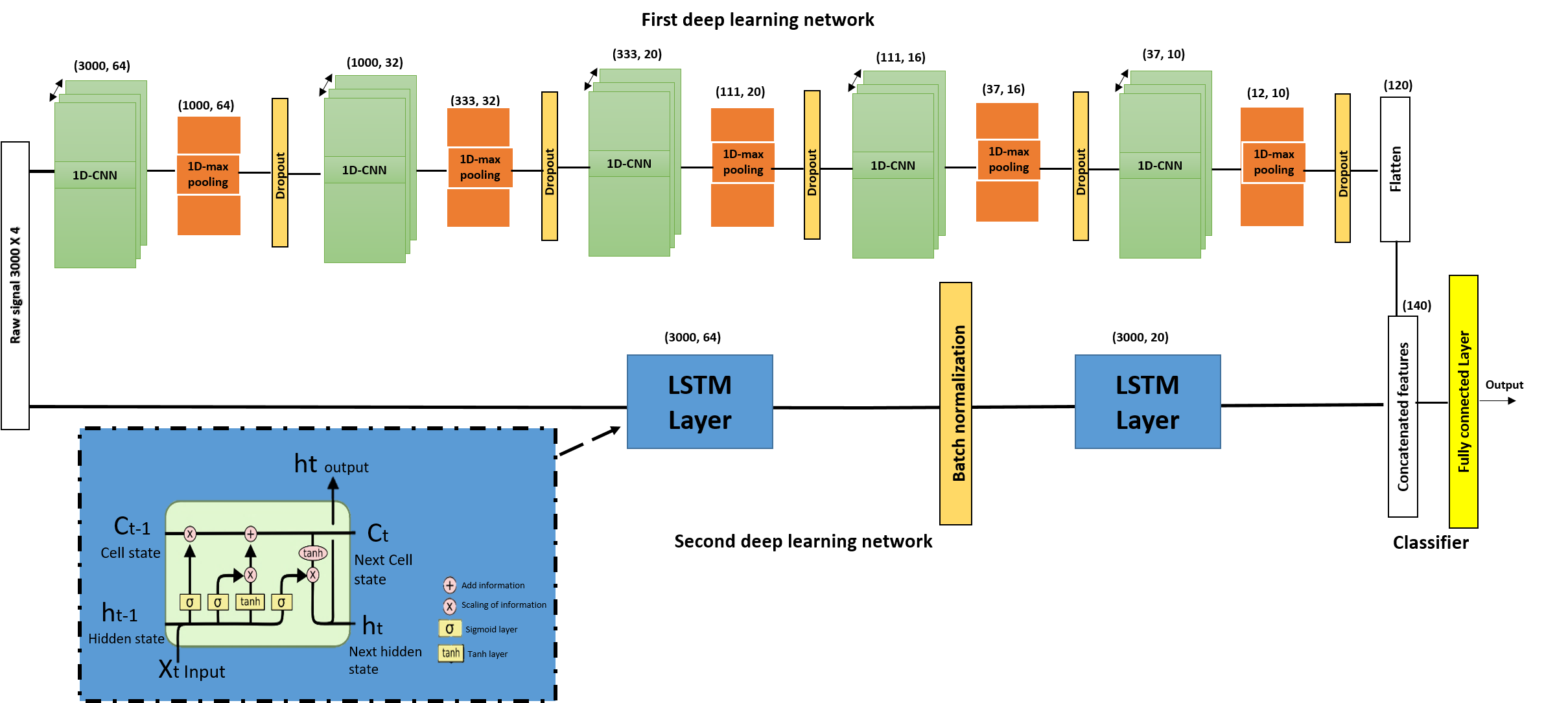}
	\caption{The detailed architecture of the proposed SSNet consists of two main deep learning networks using Sleep-EDFX dataset size. The first deep learning network is based on 1D-CNN layers with a number of feature maps are 64,32,20,16,10, respectively. Max-pooling layers with size 3 are added after each 1D-CNN layer. The second deep learning network is based on LSTM layers with sizes 64 and 20, respectively.  The classifier at the end concatenates the extracted features and predicts the final outputs by using a fully connected layer with softmax.}
	\label{FIG:11}
	\end{center}
\end{figure*}

Our proposed architecture consists of two main deep learning networks as shown in Figure \ref{FIG:11}. The combination of EEG, EMG, and EOG signals is fed into the first and second deep learning networks. The first deep learning network is based on a CNN model, while the second deep learning network is based on the LSTM model. CNN model learns filters to extract time-invariant features from the raw signals while LSTM model learns long term dependencies from the input sequences of the previous sleep stage segments. For the first and second deep learning networks, we set the sizes of CNN and LSTM to be small to select only the important features from the raw signals. The selected features produced from the first and second deep learning networks are concatenated and fed to a classifier which is a fully connected layer to predict the final results. Our architecture is designed for classifying combination of 30-second EEG, EMG and EOG segments following the standard of AASM. Table \ref{tbl1} lists the parameters of the two-deep learning networks and the classifier of SSNet.

\subsection{First deep learning network } 
We employ multi 1D-CNN layers with small filters size to extract time-invariant features from the raw signals such as specific signal patterns. The first deep learning network consists of five 1D-CNN, max-pooling and dropout layers. Each 1D-CNN layer contains \textit{Kernels} filters, which are used to extract features from the raw signal in the form of a feature map \cite{almutairi2021detection2}. The CNN layer is presented in Eq.~(\ref{Eqcnn})\cite{almutairi2021detection}:
\begin{equation}
CNN = W^l_k x^l_{i,j}+b^l_k
\label{Eqcnn}
\end{equation}

where $W$ is the weight vector, $b$ is bias, and $x$ is the raw signals. While $l$ is the layer, and $(i,j)$ is the location of the feature value in the $k$th feature map. \\

The feature maps are produced by convolving the inputs with filters and ReLU which is represented in Eq.~(\ref{Eq9})\cite{almutairi2021detection}:
\setlength{\abovedisplayskip}{4pt}
\begin{equation}
ReLU=\left\{
\begin{array}{@{}ll@{}}
0, & \text {for } x<0 \\
x, & \text{for } x=>0
\end{array}\right.
\label{Eq9} 
\end{equation}
where $x$ is the raw signal. \\

Each 1D-CNN layer is followed by 1D-max pooling to reduce feature size and the computational cost of the architecture. Then, we add a dropout layer to prevent overfitting during the training. We repeat the same order of previous layers (1D-CNN, max-pooling and dropout) four times with different parameters as presented in Table \ref{tbl1} . After that, we add a flatten layer to convert the features produced from the last 1D-max-pooling layer to a single long feature vector. The total number of features produced from the first deep learning network is 120. 

\begin{table*}[]
\begin{center}
		\caption{The detailed feature map of all layers of SSNet.}\label{tbl1}
 \begin{adjustbox}{width=1.15\columnwidth,center}
		\begin{tabular}{llccccc}
			\hline
			\textbf{Deep learning network} & \textbf{Layer name} & \textbf{feature map}& \textbf{Kernal size} & \textbf{Stride}& \textbf{Activation} \\
			\hline
			\multirow{4}{*}{\textbf{First}}
			& 1D-CNN & 	64 & 5& Same &ReLU \\ 				
			& 1D-Maxpooling & 3 &&&  \\
			& Dropout & 0.02 &   &&\\
			
			& 1D-CNN & 	32 &  3&Same&ReLU \\ 	
			& 1D-Maxpooling & 3 &&&  \\
			& Dropout & 0.02 & &&  \\
			
			& 1D-CNN & 	20 & 2& Same&ReLU \\ 	
			& 1D-Maxpooling & 3& & & \\
			& Dropout & 0.02 &&&   \\	

			& 1D-CNN & 	16 &  8&Same&ReLU \\ 	
			& 1D-Maxpooling & 3 &&&  \\
			& Dropout & 0.02 &&&   \\	
			
			& 1D-CNN & 	10 & 3&Same& ReLU \\ 	
			& 1D-Maxpooling & 3 && & \\
			& Dropout & 0.02 &  & &\\	
			& Flatten & &  & &\\				
						\hline 
												
			\multirow{4}{*}{\textbf{Second}}				
			& LSTM & 	64 & & & \\ 	
			& Recurrent dropout &0.02 & & & \\
			& BatchNormalization & &&& \\		
			& LSTM & 20& &&  \\
			& Recurrent dropout & 	0.02 & &&  \\ 	
			\hline 
			
			\multirow{3}{*}{\textbf{Classifier}}				
			& Concatenated & & && \\ 	
			& Fully connected & 3,5 &&& Softmax \\
			\hline
\end{tabular}   \end{adjustbox}
			\end{center}
\end{table*}

\subsection{Second deep learning network } 
We apply two layers of LSTM networks to capture temporal features from previous input sequences such as sleep scoring rules \cite{iber2007aasm}. For instance, sleep experts determine segments as W stage when alpha activity appears in the occipital region with more than 50\% of the segment. In this case, LSTM network can learn long term dependencies from the previous sleep stages segments to remember that it has seen W stage, and score segments as W stage if it still detects characteristics of W stage. 

LSTM network can learn long term dependencies through three gate layers: the Input gate layer, Forget gate layer and Output gate layer. The mathematical representation of Forget gate layer $f_t$ is presented in Eq.(\ref{Eqqff}), which uses a sigmoid layer to exclude some information from the cell state $C_t$. The Input gate layer $i_t$ decides to store new information in the $C_t$ by two steps. The first step, as presented in Eq~(\ref{Eqf}), is a sigmoid layer determining which values will be updated. Second step, as presented in Eq~(\ref{Eqff}), which is a $tanh$ layer creating a new candidate values ${C^\sim_t}$ that will be added to the $C_t$. Then, the previous two steps will update the old $C_{t-1}$ as presented in Eq.~(\ref{Eqfff}). The Output gate layer $h_t$ produces the output from two steps: first step, a sigmoid layer $o_t$ is used to filter the information in the $C_t$ as presented in Eq.~(\ref{Eqo}), second step, a $tanh$ layer is used to normalize the values in the $C_t$ between 1 and -1 and multiply the result from $o_t$ with a $tanh$ layer as presented in Eq.~(\ref{Eqo1}. 

\setlength{\abovedisplayskip}{4pt}
\setlength{\belowdisplayskip}{6pt}
\begin{flalign}
& f_t =\sigma(W.[h_{t-1},x_t]+b)
\label{Eqqff}\\
& i_t =\sigma(W.[h_{t-1},x_t]+b)
\label{Eqf} \\
& C^\sim_t =tanh(W.[h_{t-1},x_t]+b)
\label{Eqff}\\
& C_t =f_t*C_{t-1}+i_t*C^\sim_t
\label{Eqfff}\\
& o_t =\sigma(W.[h_{t-1},x_t]+b)
\label{Eqo}\\
& h_t = o_t * tanh(C_t)
\label{Eqo1}
\end{flalign}
where $ h_{t-1}$ is the hidden units and $x_t$ is the input feature at the time $t$, while $W$ is the weight of the inputs and $b$ is the bias. $\sigma$ is the non-linear hyperbolic function.\\

The first LSTM network is followed by batch-normalization layer to speed up the training phase. The size of the two LSTM networks are 64 and 20, respectively. The total number of features produced from the second deep learning network is 20.

\subsection{Classifier}
 We concatenate all the selected features extracted by the first and second deep learning networks, resulting in 140 features. From this step, our model enables to classify the combination of time-invariant features extracted from the CNN and the temporal features learned from the previous input sequences in LSTM networks. We add a fully connected layer with a softmax to predict the final classification results. We train our model to classify the segments into three classes: W, NREM, and REM. Then, we repeat the experiment to classify the segments into five classes: W, N1, N2, N3, and REM. 

\section{Performance Metrics}
We evaluated the performance of SSNet using machine learning metrics such as Sensitivity (SE) or Recall, Accuracy (ACC), F1 score, Specificity (SP) and Kappa coefficient. Kappa coefficient is an appropriate performance metric for assessing classification performance on an imbalanced dataset \cite{timotius2010arithmetic}. These metrics are calculated as: 

\begin{flalign}
&\text{SE}=\frac{TP}{TP+FN}
\label{eq:2}\\
&\text{SP}=\frac{TN}{TN+FP}
\label{eq:3}\\
&\text{ACC}=\frac{TN+TP}{N}
\label{eq:4}\\
&\text{Precision}=\frac{TP}{TP+FP}
\label{eq:6}\\
&\text{F1}=\frac{2( SE \times Precision)} {SE + Precision}
\label{eq:7}\\
&\text{Kappa}= \textstyle \frac{2( TN \times TP- FP \times FN)} {(TN+FN)\times (FN+TP) + (FP+TP)\times (TN+FP)}
\label{eq:8}
\end{flalign}

where TP refers to True Positive segments, TN refers to True Negative segments, FP refers to False Positives segments, and N refers to the total number of segments.

\section{Results}

We conduct experiments using Sleep EDFX dataset for three sleep stage classification to find the best input sources. We train our proposed model with single-channel of EEG: FPz-Cz, single-channel of EEG: Pz-Oz, single-channel EMG, single-channel EOG, a combination of the two channels of EEG signal and single-channel EOG, and a combination of the two channels of EEG signal and single-channel EMG. The detailed results of the performance of the proposed SSNet with single-channels of EEG, EMG and EOG signals, and the combination of signals of EEG+EOG, and EEG+EMG for the three sleep stage classification using Sleep-EDFX dataset are presented in Table \ref{allsignal}. It can be observed that the highest results of accuracy and kappa are obtained with the combination of two channels of EEG: FPz-Cz and Pz-Oz, and EMG signals, and with the same two channels of EEG and EOG signals. The average accuracy and Kappa are 95.46\% and 89.70\% respectively with the combination of two channels of EEG and EMG signals. Similarly, the average accuracy and Kappa are 95.65\%, and 90.12\% respectively for the combination of two channels of EEG and EOG signals.\\

\begin{figure*}[]
	\begin{center}
	\includegraphics[scale=.70]{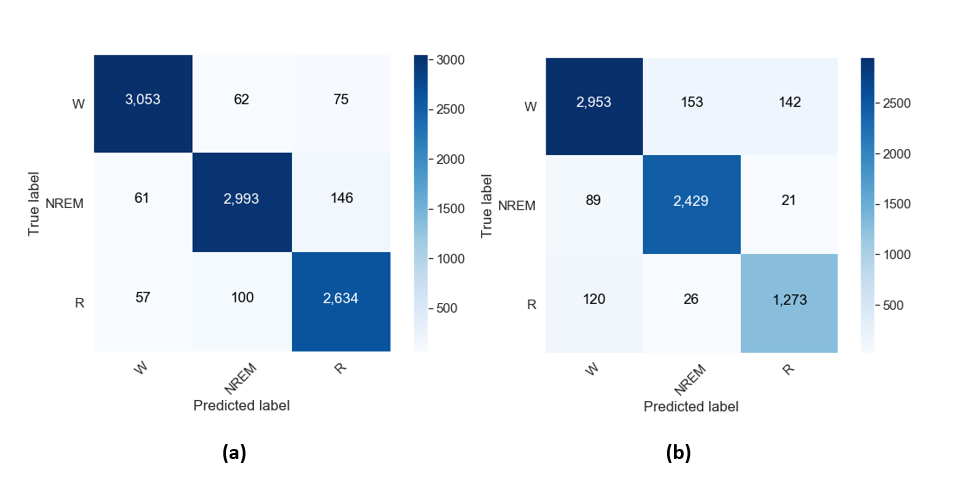}
	\caption{Confusion matrix of the three sleep stage classification
using: (a) Sleep-EDFX dataset and (b) ISRUC-sleep dataset}
	\label{FIG:confiusion 1}
		\end{center}
\end{figure*}

From the observation of results presented in Table 4, we select the two channels of EEG with EMG and EOG signals as input sources. We train our model on both the datasets: ISRUC-Sleep and Sleep-EDFX. The detailed results of the performance of the proposed SSNet for the three sleep stage classification with the combination of EEG, EMG and EOG signals are presented in Table \ref{ttbl222} and the confusion matrix is presented in Figure \ref{FIG:confiusion 1}. The average accuracy, sensitivity, specificity, F1 score and Kappa using ISRUC-Sleep dataset are 94.90\%, 92.00\%, 96.02\%, 91.90\% and 90.34\%, respectively while, the average accuracy, sensitivity, specificity, F1 score, and Kappa using Sleep-EDFX dataset are 96.36\%, 94.53\%, 97.28\%, 94.49\%, and 93.40\%, respectively. Kappa results of the proposed model with ISRUC-Sleep dataset range from 87.98\% to 98.88\%, while kappa results with Sleep-EDFX dataset range from 92.08\% to 95.15\%.\\

\begin{figure*}[]
	\begin{center}
	\includegraphics[scale=.70]{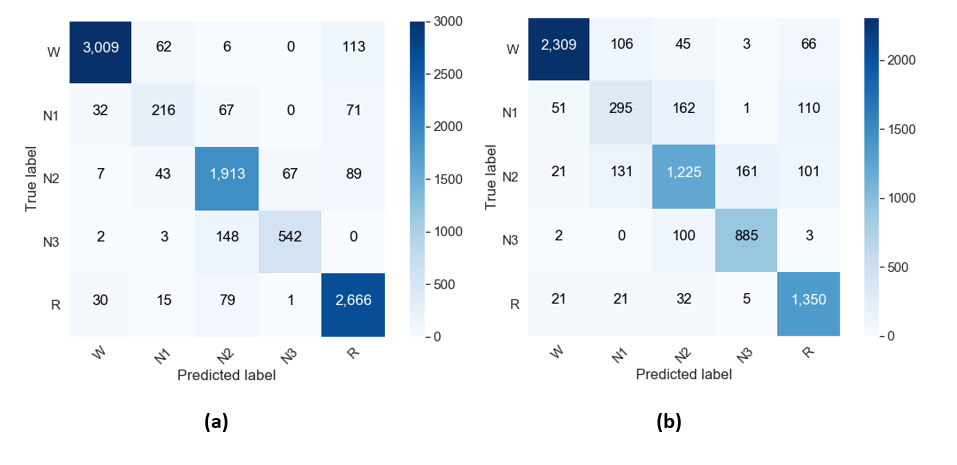}
	\caption{Confusion matrix of the five sleep stage classification
using: (a) Sleep-EDFX dataset and (b) ISRUC-sleep dataset}
	\label{FIG:confiusion 2}
		\end{center}
\end{figure*}

The detailed results of the performance of the proposed SSNet for the five sleep stage classification are presented in Table \ref{t3bl222} and confusion matrix is shown in Figure \ref{FIG:confiusion 2}. The average accuracy, sensitivity, specificity, F1 score and Kappa using ISRUC-Sleep dataset are found to be 93.69\%, 79.51\%, 96.10\%, 79.05\%, and 77.31\% respectively. For Sleep-EDFX dataset, the average accuracy, sensitivity, specificity, F1 score, and Kappa are found to be 96.57\%, 82.81\%, 97.89\%, 84\%, and 83.05\%, respectively. It can be seen that the lowest results of Kappa are obtained for N1 class with 48.24\% on ISRUC-Sleep dataset and 58.77\% on Sleep-EDFX dataset, while Kappa results for other classes are significantly better with the range from 73.55\% to 95.15\%. 

\begin{table*}[]
	\begin{center}
 \caption{The classification results of single-channel and combination signals for each class and average using SSNet for three sleep stage classification.}\label{allsignal}
 \begin{adjustbox}{width=1.10\columnwidth,center}

\begin{tabular}{|lllllllllllll|}
\hline
\multicolumn{13}{|c|}{\textbf{Sleep-EDF Expanded Dataset}}                                                                                                                                                   \\ \hline
\multicolumn{3}{|c||}{\textbf{EEG(FPz-Cz)}}                                                                     & \multicolumn{2}{c||}{\textbf{EEG(Pz-Oz)}}                                & \multicolumn{2}{c||}{\textbf{EMG}}                                       & \multicolumn{2}{l||}{\textbf{EOG}}                                       & \multicolumn{2}{c||}{\textbf{2EE+EMG}}                                   & \multicolumn{2}{c|}{\textbf{2EEG+EOG}}             \\ \hline \hline
\multicolumn{1}{|l|}{\textbf{Class}} & \multicolumn{1}{l|}{\textbf{ACC}} & \multicolumn{1}{l||}{\textbf{Kappa}} & \multicolumn{1}{l|}{\textbf{ACC}} & \multicolumn{1}{l||}{\textbf{Kappa}} & \multicolumn{1}{l|}{\textbf{ACC}} & \multicolumn{1}{l||}{\textbf{Kappa}} & \multicolumn{1}{l|}{\textbf{ACC}} & \multicolumn{1}{l||}{\textbf{Kappa}} & \multicolumn{1}{l|}{\textbf{ACC}} & \multicolumn{1}{l||}{\textbf{Kappa}} & \multicolumn{1}{l|}{\textbf{ACC}} & \textbf{Kappa} \\ \hline
\multicolumn{1}{|l|}{W}              & \multicolumn{1}{l|}{95.93}        & \multicolumn{1}{l||}{90.94}          & \multicolumn{1}{l|}{95.00}        & \multicolumn{1}{l||}{89.02}          & \multicolumn{1}{l|}{87.58}             & \multicolumn{1}{l||}{71.54}               & \multicolumn{1}{l|}{94.30}             & \multicolumn{1}{l||}{85.67}               & \multicolumn{1}{l|}{96.89}        & \multicolumn{1}{l||}{93.13}          & \multicolumn{1}{l|}{\textbf{97.08}}        & \textbf{93.54}          \\ \hline
\multicolumn{1}{|l|}{NREM}           & \multicolumn{1}{l|}{94.36}        & \multicolumn{1}{l||}{87.60}          & \multicolumn{1}{l|}{92.63}        & \multicolumn{1}{l||}{83.62}          & \multicolumn{1}{l|}{87.56}             & \multicolumn{1}{l||}{71.88}               & \multicolumn{1}{l|}{93.84}             & \multicolumn{1}{l||}{87.04}               & \multicolumn{1}{l|}{95.21}        & \multicolumn{1}{l||}{89.56}          & \multicolumn{1}{l|}{95.32}        & 89.75          \\ \hline
\multicolumn{1}{|l|}{R}              & \multicolumn{1}{l|}{92.41}        & \multicolumn{1}{l||}{82.33}          & \multicolumn{1}{l|}{90.66}        & \multicolumn{1}{l||}{78.09}          & \multicolumn{1}{l|}{82.24}             & \multicolumn{1}{l||}{61.43}               & \multicolumn{1}{l|}{92.04}             & \multicolumn{1}{l||}{82.23}               & \multicolumn{1}{l|}{94.29}        & \multicolumn{1}{l||}{86.41}          & \multicolumn{1}{l|}{94.54}        & 87.08          \\ \hline
\multicolumn{1}{|l|}{\textit{Average}}        & \multicolumn{1}{l|}{\textit{94.24}}        & \multicolumn{1}{l||}{\textit{86.96} }         & \multicolumn{1}{l|}{\textit{92.76}}        & \multicolumn{1}{l||}{\textit{83.58}}          & \multicolumn{1}{l|}{\textit{85.79} }            & \multicolumn{1}{l||}{\textit{68.28}}               & \multicolumn{1}{l|}{\textit{93.39}}             & \multicolumn{1}{l||}{\textit{84.98}}               & \multicolumn{1}{l|}{\textit{95.46}}        & \multicolumn{1}{l||}{\textit{89.70} }         & \multicolumn{1}{l|}{\textit{95.65}}        & \textit{90.12}          \\ \hline

\end{tabular}\end{adjustbox}
\footnotesize{ Bold represnts the best results.
ACC = Accuracy, Kappa= Kappa coefficient.} 
	\end{center}
\end{table*}

\begin{table*}[]
\begin{center}
 \caption{The classification results for each class and average using SSNet for three sleep stage classification.}\label{ttbl222}
  \begin{adjustbox}{width=1.15\columnwidth,center}

\begin{tabular}{|l|lllll|lllll|}
\hline
\textbf{}        & \multicolumn{5}{l|}{\textbf{ISRUC-Sleep Dataset}}                                                                                                                      & \multicolumn{5}{l|}{\textbf{Sleep-EDF Expanded Dataset}}                                                                                                               \\ \hline
\textbf{Class}   & \multicolumn{1}{c|}{\textbf{ACC}}   & \multicolumn{1}{c|}{\textbf{SE}}    & \multicolumn{1}{c|}{\textbf{SP}}    & \multicolumn{1}{c|}{\textbf{F1}}    & \textbf{Kappa} & \multicolumn{1}{c|}{\textbf{ACC}}   & \multicolumn{1}{c|}{\textbf{SE}}    & \multicolumn{1}{c|}{\textbf{SP}}    & \multicolumn{1}{c|}{\textbf{F1}}    & \textbf{Kappa} \\ \hline
W                & \multicolumn{1}{l|}{93.00}          & \multicolumn{1}{l|}{90.91}          & \multicolumn{1}{l|}{94.72}          & \multicolumn{1}{l|}{92.13}          & \textbf{98.88}          & \multicolumn{1}{l|}{97.22}          & \multicolumn{1}{l|}{95.70}          & \multicolumn{1}{l|}{98.03}          & \multicolumn{1}{l|}{95.99}          & \textbf{95.15}          \\ \hline
NREM             & \multicolumn{1}{l|}{95.98}          & \multicolumn{1}{l|}{95.66}          & \multicolumn{1}{l|}{96.16}          & \multicolumn{1}{l|}{94.38}          & 93.16          & \multicolumn{1}{l|}{95.98}          & \multicolumn{1}{l|}{93.53}          & \multicolumn{1}{l|}{97.29}          & \multicolumn{1}{l|}{94.19}          & 92.97          \\ \hline
R                & \multicolumn{1}{l|}{95.71}          & \multicolumn{1}{l|}{89.71}          & \multicolumn{1}{l|}{97.18}          & \multicolumn{1}{l|}{89.17}          & 87.98          & \multicolumn{1}{l|}{95.88}          & \multicolumn{1}{l|}{94.37}          & \multicolumn{1}{l|}{96.54}          & \multicolumn{1}{l|}{93.30}          & 92.08          \\ \hline
\textit{Average} & \multicolumn{1}{l|}{\textit{94.90}} & \multicolumn{1}{l|}{\textit{92.00}} & \multicolumn{1}{l|}{\textit{96.02}} & \multicolumn{1}{l|}{\textit{91.90}} & \textit{90.34} & \multicolumn{1}{l|}{\textit{96.36}} & \multicolumn{1}{l|}{\textit{94.53}} & \multicolumn{1}{l|}{\textit{97.28}} & \multicolumn{1}{l|}{\textit{94.49}} & \textit{93.40} \\ \hline
\end{tabular}\end{adjustbox}
\footnotesize{Bold represents the best results.
ACC = Accuracy, SE = Sensitivity, SP = Specificity, Kappa=
Kappa coefficient}.
\end{center}

\end{table*}

\begin{table*}[]
\begin{center}
\caption{The classification results for each class and average using SSNet for five sleep stage classification.}\label{t3bl222}
  \begin{adjustbox}{width=1.15\columnwidth,center}

\begin{tabular}{|l|lllll|lllll|}
\hline
\textbf{}        & \multicolumn{5}{l|}{\textbf{ISRUC-Sleep Dataset}}                                                                                                                      & \multicolumn{5}{l|}{\textbf{Sleep-EDF Expanded Dataset}}                                                                                                                                    \\ \hline
\textbf{Class}   & \multicolumn{1}{c|}{\textbf{ACC}}   & \multicolumn{1}{c|}{\textbf{SE}}    & \multicolumn{1}{c|}{\textbf{SP}}    & \multicolumn{1}{c|}{\textbf{F1}}    & \textbf{Kappa} & \multicolumn{1}{c|}{\textbf{ACC}}   & \multicolumn{1}{c|}{\textbf{SE}}    & \multicolumn{1}{c|}{\textbf{SP}}    & \multicolumn{1}{c|}{\textbf{F1}}    & \multicolumn{1}{c|}{\textbf{Kappa}} \\ \hline
W                & \multicolumn{1}{l|}{95.80}          & \multicolumn{1}{l|}{91.30}          & \multicolumn{1}{l|}{98.08}          & \multicolumn{1}{l|}{93.61}          & \textbf{92.35} & \multicolumn{1}{l|}{97.25}          & \multicolumn{1}{l|}{94.32}          & \multicolumn{1}{l|}{98.81}          & \multicolumn{1}{l|}{95.98}          & \textbf{95.15}                      \\ \hline
N1               & \multicolumn{1}{l|}{91.92}          & \multicolumn{1}{l|}{47.65}          & \multicolumn{1}{l|}{96.08}          & \multicolumn{1}{l|}{50.34}          & 48.24          & \multicolumn{1}{l|}{96.80}          & \multicolumn{1}{l|}{55.95}          & \multicolumn{1}{l|}{98.60}          & \multicolumn{1}{l|}{59.58}          & 58.77                               \\ \hline
N2               & \multicolumn{1}{l|}{89.55}          & \multicolumn{1}{l|}{74.74}          & \multicolumn{1}{l|}{93.91}          & \multicolumn{1}{l|}{76.49}          & 73.55          & \multicolumn{1}{l|}{94.48}          & \multicolumn{1}{l|}{90.27}          & \multicolumn{1}{l|}{95.75}          & \multicolumn{1}{l|}{88.31}          & 86.75                               \\ \hline
N3               & \multicolumn{1}{l|}{96.18}          & \multicolumn{1}{l|}{89.39}          & \multicolumn{1}{l|}{97.29}          & \multicolumn{1}{l|}{86.55}          & 85.52          & \multicolumn{1}{l|}{97.59}          & \multicolumn{1}{l|}{77.98}          & \multicolumn{1}{l|}{99.19}          & \multicolumn{1}{l|}{83.00}          & 82.44                               \\ \hline
R                & \multicolumn{1}{l|}{95.01}          & \multicolumn{1}{l|}{94.47}          & \multicolumn{1}{l|}{95.15}          & \multicolumn{1}{l|}{88.26}          & 86.87          & \multicolumn{1}{l|}{96.73}          & \multicolumn{1}{l|}{95.52}          & \multicolumn{1}{l|}{97.09}          & \multicolumn{1}{l|}{93.05}          & 92.21                               \\ \hline
\textit{Average} & \multicolumn{1}{l|}{\textit{93.69}} & \multicolumn{1}{l|}{\textit{79.51}} & \multicolumn{1}{l|}{\textit{96.10}} & \multicolumn{1}{l|}{\textit{79.05}} & \textit{77.31} & \multicolumn{1}{l|}{\textit{96.57}} & \multicolumn{1}{l|}{\textit{82.81}} & \multicolumn{1}{l|}{\textit{97.89}} & \multicolumn{1}{l|}{\textit{84.00}} & \textit{83.05}                      \\ \hline
\end{tabular}\end{adjustbox}\\
\footnotesize{Bold represents the best results.
ACC = Accuracy, SE = Sensitivity, SP = Specificity, Kappa=
Kappa coefficient.} 
\end{center}

\end{table*}


\begin{table*}[]
	\begin{center}
 \caption{The performance of proposed model and the state-of-the-art models for three sleep stages classification using Sleep-EDF and Sleep-EDFX datasets.}\label{tbbl11}
  \begin{adjustbox}{width=1.15\columnwidth,center}

\begin{tabular}{|l|l|l|c|l|l|l|}
\hline
\textbf{Authors}                                                                & \textbf{Dataset} & \textbf{Method}    & \textbf{Number of segments} & \textbf{Signals}                                              & \textbf{Accuracy} & \textbf{Kappa} \\ \hline
Hassan et al. \cite{hassan2017automated}                         & Sleep-EDF        & EEMD+RUSBoost      & 15,188                    & 1- EEG                                                        & 94.23             & 84.70          \\ \hline
Hassan et al. \cite{hassan2016computer}                          & Sleep-EDF        & CEEMDAN+ Bagging   & 15,188                    & 1- EEG                                                        & 94.10             & 93.00             \\ \hline
Zhu et al. \cite{zhu2014analysis}                                & Sleep-EDF        & HVG+SVM            & 14,963                    & 1- EEG                                                        & 92.60             & 87.00             \\ \hline
Sharma et al. \cite{sharma2018accurate}                          & Sleep-EDF        & Wavelet filter+SVM & 85,900                     & 1-   EEG                                                      & 92.10             & 56.80              \\ \hline
Yildirim et al. \cite{yildirim2019deep}                                                                  & Sleep-EDF        & 1D-CNN             & 15,188                    & \begin{tabular}[c]{@{}l@{}}1-EEG\\ 1-EOG \end{tabular}                                                  & 94.20             & -              \\ \hline
Yildirim et al. \cite{yildirim2019deep}                                                                  & Sleep-EDFX       & 1D-CNN             & 127,512                   & \begin{tabular}[c]{@{}l@{}}1-EEG\\ 1-EOG \end{tabular}                                                   & 94.23             & -              \\ \hline

Proposed model                                                                    & Sleep-EDFX       & SSNet       & 72,000                    & \begin{tabular}[c]{@{}l@{}}2-EEG\\ 1-EOG\\ 1-EMG\end{tabular} & \textbf{96.36}             & \textbf{93.40 }         \\ \hline
\end{tabular} \end{adjustbox}\\
\footnotesize{Bold represents the best results.}
     	\end{center}

\end{table*}

\begin{table*}[]
	\begin{center}

 \caption{The performance of our proposed model and the state-of-the-art models for five sleep stages classification using Sleep-EDF and Sleep-EDFX datasets.}\label{tbl111}
   \begin{adjustbox}{width=1.15\columnwidth,center}

\begin{tabular}{|l|l|l|c|l|l|l|}
\hline
\textbf{Authors}                                                                & \textbf{Dataset} & \textbf{Method}        & \textbf{Number of segments} & \textbf{Signals}                                              & \textbf{Accuracy} & \textbf{Kappa} \\ \hline
Hassan et al.\cite{hassan2017automated}                         & Sleep-EDF        & EEMD+RUSBoost          & 15,188                    & 1- EEG                                                        & 83.49             & \textbf{84.05}        \\ \hline
Hassan et al. \cite{hassan2016computer}                           & Sleep-EDF        & CEEMDAN+ Bagging       & 15,188                    & 1- EEG                                                        & 90.69             & \textbf{89.00}           \\ \hline
Zhu et al. \cite{zhu2014analysis}                                & Sleep-EDF        & HVG+SVM                & 14,963                    & 1- EEG                                                        & 88.90             & 83.00           \\ \hline
Sharma et al. \cite{sharma2018accurate}                          & Sleep-EDF        & Wavelet filter+SVM     & 85,900                    & 1-   EEG                                                      & 91.50              & 58.81          \\ \hline
Yildirim et al. \cite{yildirim2019deep}                                                                  & Sleep-EDF        & 1D-CNN                 &15,188                   &\begin{tabular}[c]{@{}l@{}}1-EEG\\ 1-EOG \end{tabular}                                                   & 91.22             & -              \\ \hline
Rahman et al.\cite{rahman2018sleep}                                                                     & Sleep-EDF        & DWT+SVM                & 15,188                     & 1- EOG                                                        & 90.20              & -               \\ \hline
Nguyen et al. \cite{rajbhandari2021novel}                                                                    & Sleep-EDFX        & 1D-CNN                 & 3,000                     & 1- EGG                                                        & 87.67             & -               \\ \hline

Yildirim et al. \cite{yildirim2019deep}                                                                  & Sleep-EDFX       & 1D-CNN                 & 127,512                   & \begin{tabular}[c]{@{}l@{}}1-EEG\\ 1-EOG \end{tabular}                                                    & 90.98             & -              \\ \hline
Satapathy et al. \cite{satapathy2021machine}                     & Sleep-EDFX       & Statistic features+ RF & 15,139                    & 1-EEG                                                         & 92.79             & \textbf{88.00}            \\ \hline
Rahman et al. \cite{rahman2018sleep}                             & Sleep-EDFX       & DWT+SVM                & 54,587                   & 1- EOG                                                        & 91.70             & -              \\ \hline
Zhu et al. \cite{zhu2020convolution}                                                                  & Sleep-EDFX       & Attention CNN             & 42,269                  & 1-EEG                                                  & 82.80             & 77.34              \\ \hline
Proposed model                                                                    & Sleep-EDFX       & SSNet           & 72,000                    &  \begin{tabular}[c]{@{}l@{}}2-EEG\\ 1-EOG\\ 1-EMG\end{tabular} & \textbf{96.57}             & 83.05         \\ \hline
\end{tabular}   \end{adjustbox}
\footnotesize{Bold represents the best results.} 
	\end{center}
\end{table*}

\begin{table*}[]
	\begin{center}
 \caption{The performance of proposed model and the state-of-the-art models for three and five sleep stages classification using ISRUC-Sleep dataset.}\label{tbl11}
    \begin{adjustbox}{width=1.15\columnwidth,center}

\begin{tabular}{|l|l|l|c|l|l|l|}
\hline
\textbf{Authors} & \textbf{Classes} & \textbf{Method} & \textbf{Number of segments} & \textbf{Signals}                                              & \textbf{Accuracy} & \textbf{Kappa} \\ \hline
Nguyen et al. \cite{rajbhandari2021novel}      & 5                & 1D-CNN          & 3,000           & 1- EEG                                                        &86.76            & -              \\ \hline
Rahman et al.\cite{rahman2018sleep}      & 5                & DWT+SVM         & 9,001           & 1- EOG                                                        & 86.00                & -              \\ \hline
Proposed model     & 3                & SSNet    & 56,515          & \begin{tabular}[c]{@{}l@{}}2-EEG\\ 1-EOG\\ 2-EMG\end{tabular} & \textbf{94.90 }            & \textbf{90.34 }        \\ \hline
proposed model     & 5                & SSNet    & 56,515          & \begin{tabular}[c]{@{}l@{}}2-EEG\\ 1-EOG\\ 2-EMG\end{tabular} & \textbf{93.96}             & \textbf{77.31 }       \\ \hline
\end{tabular}   \end{adjustbox}
\\
\footnotesize{Bold represents the best results.} 
 \end{center}     	
\end{table*}

\begin{table}[]
	\begin{center}
\caption{Comparison of REM detection in terms of precision and recall for five sleep stage classification by the proposed model and the state-of-the-art methods.}\label{tbl44}
\begin{tabular}{|l|l|l|}
\hline
\multirow{2}{*}{\textbf{Research studies}}                & \multirow{2}{*}{\textbf{Precision}} & \multirow{2}{*}{\textbf{Recall}} \\
                                                          &                                     &                                  \\ \hline
Hassan et al. \cite{hassan2016computer} & 80.17                               & 80.86                            \\ \hline
Sharma et al. \cite{sharma2018accurate}  & 46.87                               & 36.45                            \\ \hline
Zhu et al. \cite{zhu2014analysis}        & 76.21                               & 72.85                            \\ \hline
Zhu et al. \cite{zhu2020convolution}     & 82                                  & 84.60                            \\ \hline
Rahman et al. \cite{rahman2018sleep}     & -                                   & 84.70                            \\ \hline
Proposed model                                            & \textbf{90.71}                      & \textbf{95.52}                   \\ \hline
\end{tabular}\\
\footnotesize{Bold represents the best results.}
	\end{center}
\end{table}

\section{Discussion}
We compare the results obtained by our proposed model with the most relevant state-of-the-art models, which address the three and five sleep stage classifications. We include all the state-of-the-art studies that use a splitting strategy of the datasets in which the segments from the same patient recording are split into training and testing sets. Many studies propose feature engineering methods and machine learning models for the three and five sleep stage classifications, while a few studies use deep learning models without any feature engineering methods.  \\

Table \ref{tbbl11} presents a comparison of accuracy and kappa between our proposed model and the state-of-the-art models using Sleep-EDF and Sleep-EDFX datasets. Most of the state-of-the-art studies did not provide the other evaluation metrics. 
The total number of segments of sleep stages obtained from Sleep-EDFX dataset is 72,000 segments. We achieved an accuracy of 96.36\%, approximately 3\% higher than the existing state-of-the-art result. It can also be observed from kappa results that our proposed model is found to be better (93.40\%) than the stat-of-the-art models. Hassan et al. \cite{hassan2016computer} achieved 93\% of kappa which is approximately the same as our kappa result using a small number of segments in their study (15,188 segments). Overall, we conclude that the performance of our proposed model for classification of the three sleep stages achieved promising results and setting new state-of-the-art result.

Table \ref{tbl111} presents the performance of our proposed model and state-of-the-art models for five sleep stage classification using Sleep-EDF and Sleep-EDFX datasets. 
We obtained an accuracy of 96.57\%, approximately 5\% higher compared to the other state-of-the-art models. Comparing the Kappa results, Hassan et al. \cite{hassan2017automated, hassan2016computer} and Satapathy et al. \cite{satapathy2021machine} achieved higher kappa results of 84\%, 89\% and 88\%, respectively with comparatively smaller number of segments as compared to our study as presented in  Table~\ref{tbl111}. In addition, these studies require feature engineering methods for extracting features from PSG signals to classify them by traditional machine learning models. These models may cause overfitting if high dimensional PSG signals are used to train their models as reported in \cite{faust2018deep, faust2019review}. In addition, it is also reported that use of feature engineering methods to convert PSG signals to lower-dimensional feature vectors may result in information loss \cite{faust2019review}. 
Therefore, we believe that our proposed deep learning model is appropriate to use for five stage classification as it provides good performance in large dataset and more generalised.  

Table \ref{tbl11} presents the performance of our proposed model and state-of-the-art models for three and five sleep stage classifications using ISRUC-Sleep dataset. 
We used 56,515 segments of sleep stages obtained from ISRUC-Sleep dataset. We obtained the accuracies of the three and five sleep stage classification to be 94.90\% and 93.96\%, respectively. These results are the higher as compared to the state-of-the-art results by 7\%. The state-of-the-art studies \cite{rajbhandari2021novel, rahman2018sleep} did not provide Kappa results in their article to be compared.\\

It can be observed from Table \ref{ttbl222} and \ref{t3bl222} that our proposed model can classify the combination of signals with different sampling frequencies. Our model achieved good performance using ISRUC-Sleep dataset (200 Hz) as well as Sleep-EDFX dataset (100 Hz). However, it can also be observed from Table \ref{t3bl222} that N1 class did not performed well as compared to other classes. This low performance in identifying N1 class can be attributed to the common transition characteristics of W stage to other sleep stages, making it harder to distinguish as reported by Khalighi et al.~\cite{khalighi2016isruc}.  \\

Detection of REM stage is essential for diagnosing sleep disorders including narcolepsy and REM behaviour disorder \cite{iranzo2006rapid}. Previous research studies suggested that EOG and EMG signals may provide discriminative features to detect REM stage from the other sleep stages \cite{loh2020automated}. Therefore, we compare the performance of our model in detecting REM stage using evaluation metrics of precision and recall with the state-of-the-art methods. Few studies \cite{hassan2016computer, sharma2018accurate, zhu2014analysis, zhu2020convolution, rahman2018sleep} provided their confusion
matrix which helped us to calculate precision and recall of REM stage for those studies. Table \ref{tbl44} presents REM stage precision and recall in the classification of five classes of the proposed model and the existing state-of-the-art studies. It can be observed that our proposed model demonstrates that the combination of EEG, EMG and EOG improves the detection of REM stage which is found to be 90.71\% precision and 95\% recall which is higher as compared to the state-of-the-art studies. \\  

SSNet has several advantages over the existing studies. Firstly, it uses multi-channels of EEG, EMG and EOG signals, which help to improve the classification results. Secondly, we propose CNN and LSTM, which work alongside efficiently to extract features automatically instead of using complicated feature engineering methods. Thirdly, we thoroughly tested our proposed model by using two popular datasets for the classification of three and five sleep stage classes. The results show that our proposed network perform better for both classification problem.

\section{Conclusion} 
In this paper, we introduced a novel deep learning model, called SSNet. Our proposed model classified 30-second segments of a combination of EEG, EOG and EMG signals to classify three and five sleep stages. SSNet contains two deep learning networks. The first deep learning network is composed of CNN, while the second deep learning network is composed of LSTM networks. The extracted features of both networks are concatenated and passed to the fully connected layer for classification. The results demonstrated that the combination of EEG, EOG and EMG signals contributed significantly in improving the performance of classification using our proposed model. The accuracy and Kappa achieved by SSNet for three sleep stage classification were 96.36\%, and 91.81\%, respectively  while the accuracy and Kappa achieved by SSNet for five sleep stage classification were 96.57\%, and 87.43\%, respectively. The limitation of our work is the low performance of our proposed model for the detection of N1 class. 

\bibliography{sn-article}

\end{document}